# Chiral Superfluid Helium-3 in the Quasi-Two-Dimensional Limit

Petri J. Heikkinen[*], Lev V. Levitin, Xavier Rojas, Angadjit Singh, Nathan Eng, Andrew Casey, and John Saunders[†]

*Department of Physics, Royal Holloway, University of London, Egham TW20 0EX, Surrey, United Kingdom*

Anton Vorontsov

*Department of Physics, Montana State University, Bozeman, Montana 59717, USA*

Nikolay Zhelev[‡], Abhilash Thanniyil Sebastian[§], and Jeevak M. Parpia

*Department of Physics, Cornell University, Ithaca, New York 14853, USA*

Anisotropic pair breaking close to surfaces favors the chiral $A$ phase of the superfluid $^3$He over the time-reversal invariant $B$ phase. Confining the superfluid $^3$He into a cavity of height $D$ of the order of the Cooper pair size characterized by the coherence length $\xi_0$—ranging between 16 nm (34 bar) and 77 nm (0 bar)—extends the surface effects over the whole sample volume, thus allowing stabilization of the $A$ phase at pressures $P$ and temperatures $T$ where otherwise the $B$ phase would be stable. In this Letter, the surfaces of such a confined sample are covered with a superfluid $^4$He film to create specular quasiparticle scattering boundary conditions, preventing the suppression of the superfluid order parameter. We show that the chiral $A$ phase is the stable superfluid phase under strong confinement over the full $P$-$T$ phase diagram down to a quasi-two-dimensional limit $D/\xi_0 = 1$, where $D = 80$ nm. The planar phase, which is degenerate with the chiral $A$ phase in the weak-coupling limit, is not observed. The gap inferred from measurements over the wide pressure range from 0.2 to 21.0 bar leads to an empirical ansatz for temperature-dependent strong-coupling effects. We discuss how these results pave the way for the realization of the fully gapped two-dimensional $p_x + ip_y$ superfluid under more extreme confinement.

Chiral superconductivity is a rare phenomenon with nontrivial topology, predicted to result in several exotic properties [1]: macroscopic angular momentum [2–4]; anomalous quantum Hall effect [2,5,6]; half-quantum vortices [7–10]; spontaneous chiral edge currents [4,11]; and chiral Majorana states bound on the edges and vortex cores [6,10,12–15], obeying non-Abelian braiding statistics required for topological quantum computation [16–19]. Various superconducting materials have been proposed to realize chiral Cooper pairing [20,21], e.g., $Sr_2RuO_4$ [22–24], $UPt_3$ [25–28], $UTe_2$ [29], $LaPt_3P$ [30], $SrPtAs$ [31,32], and $URu_2Si_2$ [33]. Although most of these materials show evidence for broken time-reversal symmetry, the actual pairing states are still under debate [1,20,21]. Conversely, the unconventional $p$-wave spin-triplet pairing in superfluid helium-3 is well-established [34–36], with one of its stable phases, $^3$He-$A$, having a directly measurable chirality [37–41]. Importantly, the coherent nuclear dipole interactions, determining the weak spin-orbit coupling in superfluid $^3$He, allow the interrogation of the nuclear spins of the paired $^3$He fermions directly by nuclear magnetic resonance (NMR) to determine the order parameter.

Bulk $^3$He is distinguished from its metal counterparts by the maximal symmetry group $SO(3)_L \times SO(3)_S \times U(1)_\phi \times T \times C \times P$ of the normal state [35,42]. Here, $SO(3)_L$ and $SO(3)_S$ denote the three-dimensional rotations in the orbital and spin spaces, respectively, and $U(1)_\phi$ is the gauge symmetry. These continuous symmetry groups are combined with the time-reversal (T), particle-hole (C), and parity (P) discrete symmetries. Consequently, the irreducible representations from which the order parameter is constructed are simple spherical harmonics of degree one. In the bulk liquid at zero magnetic field there are two stable phases with distinct broken symmetries: the chiral $A$ phase with point nodes in the energy gap and the time-reversal-invariant $B$ phase with isotropic gap. $^3$He-$B$ dominates the $P$-$T$ phase diagram with the chiral $^3$He-$A$ restricted to temperatures relatively close to the bulk superfluid transition temperature $T_{c0}$ at pressures above the polycritical point at $P_c = 21.22$ bar; see Fig. 1(a) [35].

[*]Contact author: petri.heikkinen@rhul.ac.uk
[†]Contact author: j.saunders@rhul.ac.uk
[‡]Department of Physics, University of Oregon, Eugene, Oregon 97403, USA.
[§]VTT Technical Research Centre of Finland Ltd, Espoo 02150, Finland.

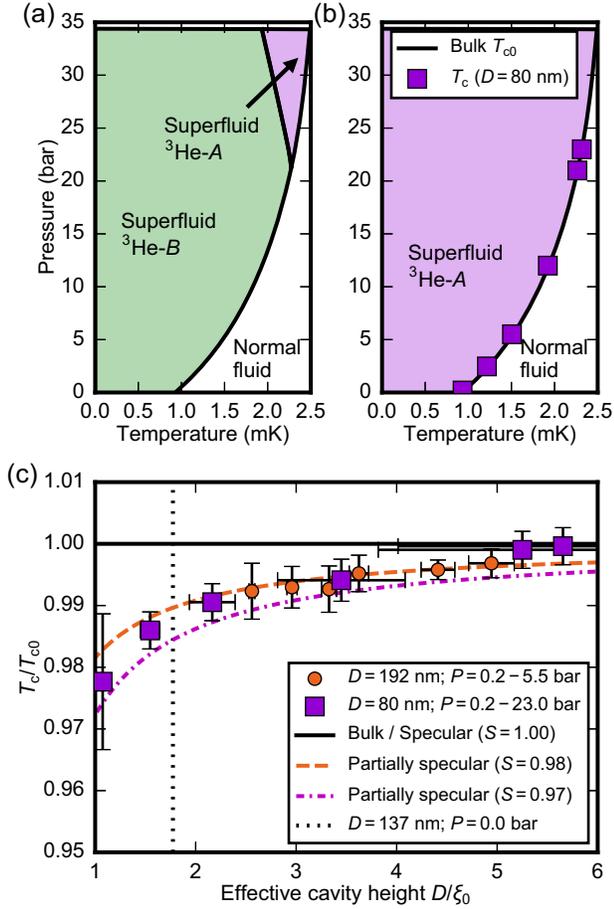

FIG. 1. Phase diagrams of bulk $^3$He with two superfluid phases (a) and of $^3$He in the $D = 80$ nm cavity, where $^3$He-$A$ is the only stable superfluid phase (b). The pressures covered in this Letter were 0.2, 2.5, 5.5, 12.0, 21.0, and 23.0 bar (violet squares). (c) The suppression of superfluid transition temperature $T_c/T_{c0}$ by confinement against the effective cavity height $D/\xi_0$ driven with geometry and pressure, with smaller values of $D/\xi_0$ corresponding to lower pressures. These universal coordinates demonstrate good agreement with the measurements in the $D = 192$ nm cavity with similar $^4$He preplating (orange circles) [43]. The suppression of $T_c$ is extremely small, consistent with calculated specularity $S > 0.97$ (dash-dotted line). The horizontal error bars encompass the cavity height distortion induced by pressure (maximally 2.6 nm/bar), as estimated from a finite-element model [43]. The vertical dotted line denotes the transport anomaly reported in a $^3$He film at $D = 137$ nm at 0 bar [44–46].

In this Letter we explore the effect of very strong anisotropic confinement on the superfluid. This breaks its rotational symmetries, leading to the predicted stability of the chiral $A$ phase over the entire $P$-$T$ phase diagram [43,57–61]. Confinement in the absence of disorder is achieved in a nanofabricated slab geometry of height $D = 80$ nm, which is comparable to the zero-temperature coherence length $\xi_0 = \hbar v_F/2\pi k_B T_{c0}$ and approaches the quasi-two-dimensional limit $D/\xi_0 = 1$ at low pressures. Here, $v_F$ is the Fermi velocity and $T_{c0}$ is the bulk superfluid transition temperature. The observed order-parameter suppression arising from pair breaking by surface scattering [57,62–65], which we tune to be specular, is minimal [43]. As a consequence, access is opened to even stronger confinement in which the size quantization across the sample can play a significant role. This potentially leads to a fully gapped two-dimensional chiral $^3$He-$A$. The absence of gap nodes in the pure 2D case eliminates the nodal low-energy quasiparticles, leaving Majorana-Weyl edge states as the only available subgap states. Moreover, tuning the cavity height is predicted to lead to analogs of the quantum Hall effect [2,5,34].

An important consideration is the relative stability of the chiral $A$ phase and the time-reversal-invariant planar phase (2D helical phase), particularly at low pressures, since these phases are degenerate in the weak-coupling limit [35]. However, in our experiment we found no evidence of any phase other than the $A$ phase. This discovery does not support the interpretation of a transport anomaly observed at a film thickness $D = 137$ nm in terms of a phase transition from $^3$He-$A$ into a different phase [44–46]. Furthermore, having access to the $A$ phase over a wide temperature and pressure range, we characterized the pressure and temperature dependence of the strong-coupling effects stabilizing it.

We report superconducting quantum interference device-amplified NMR experiments performed using the setup described in Refs. [43,61,66,67]. The sample container with atomically smooth walls (0.1 nm surface roughness) was nanofabricated out of two silicon pieces. The silicon patterning process followed Refs. [68,69]; an additional step of thermal oxidation passivated the surfaces [70,71] before fusion bonding the patterned wafer and the lid together [46,67,72]. The cavity was connected to the fill line by a set of five 2.5 mm × 10 μm × 50 μm trenches [67] providing the "bulk marker" for in situ determination of $T_{c0}$ [43]. We performed NMR in a magnetic field of approximately 30 mT perpendicular to the cavity, $\mathbf{H}_0 \parallel \hat{\mathbf{z}}$, corresponding to the Larmor frequency $f_L = 1005$ kHz. Magnetic field gradients were applied to separate the signals arising from the cavity and the bulk marker, and to suppress the signal from the mouth of the fill line.

The temperature of the heat exchanger was measured with a Pt-NMR thermometer, calibrated against the $^3$He melting curve [73]; the temperature gradient between the helium in the cavity and the Pt sensor was inferred as described in the supplementary information of Ref. [43].

To achieve essentially specular quasiparticle scattering at the boundaries [43,61,74,75], we preplated the surfaces with 100 μmol/m$^2$ of $^4$He prior to filling the container with

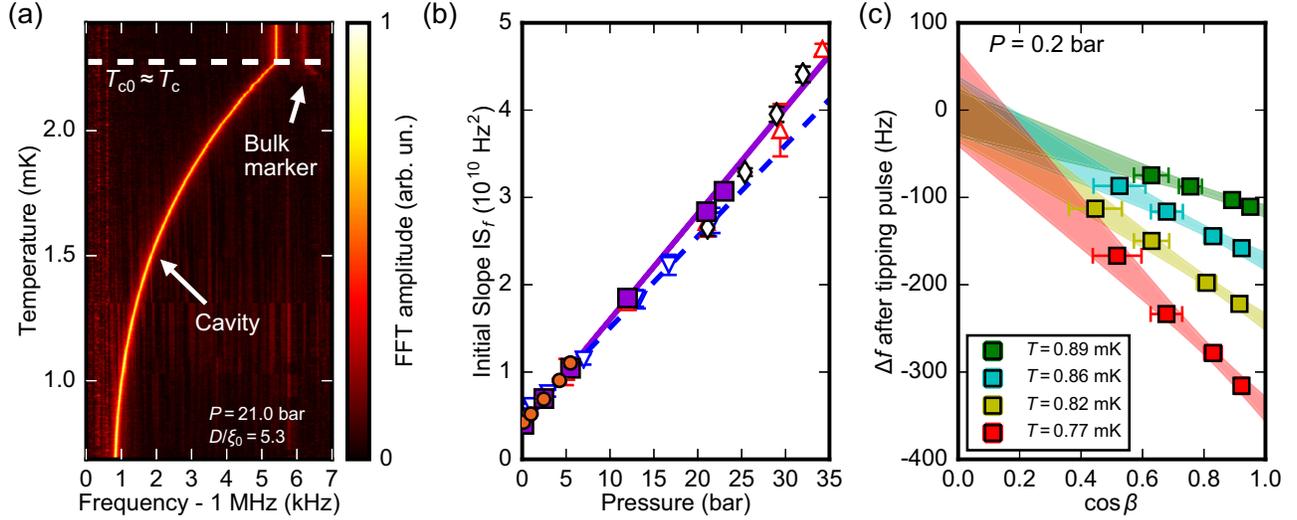

FIG. 2. NMR signatures of the superfluidity in the 80 nm high slab-shaped cavity. (a) Color map of NMR spectra arising from the cavity and the bulk marker as a function of temperature at 21.0 bar. The cavity signal has a constant magnitude and negative frequency shift below the superfluid transition temperature $T_c$, characteristic of the $^3$He-$A$ with dipole-unlocked spin-orbit orientation. The bulk marker, shifted to higher normal-state frequency by a magnetic field gradient, shows the expected positive frequency shift below the bulk superfluid transition temperature $T_{c0}$ and is lost on cooling due to broadening. The white dashed line indicates the measured superfluid transition temperatures in the cavity and in the bulk marker, which at $P = 21.0$ bar are virtually equal [see Fig. 1(c)]. (b) The initial slope of the frequency shift in the cavity in comparison with prior measurements in the $A$ phase. Closed symbols mark the values determined by us under confinement with the specular boundary condition: $D = 80$ nm (violet squares) and $D = 192$ nm (orange circles) [43]. The violet line is a linear fit to these combined data sets [46]. The high-field experiments (blue downward-pointing triangles) and the linear fit (blue dashed line) of the bulk $^3$He-$A$ initial slope are from Refs. [77–79]. The data based on stable or supercooled bulk $^3$He-$A$ are from Ref. [80] (red upward-pointing triangles) and Ref. [81] (black diamonds). To compensate for the systematic differences arising from different definitions of $IS_f$, small adjustments have been made to some of the values [43,46]. (c) The dependence of the cavity frequency shift $\Delta f$ on the tipping angle $\beta$ is consistent with $\Delta f(\beta) \propto -\cos\beta$ (colored bands), a signature of the $A$ phase with $\hat{\mathbf{d}} \perp \hat{\mathbf{l}} \| \mathbf{H}_0$. The range of $\beta$ was limited to $\beta < 60°$ by anomalous NMR heating of confined helium [82]. To mitigate the effect the measurements employed the "pulse-antipulse" technique to vary $\beta$ at a constant level of heating [83]. The legend quotes the temperature of confined helium inferred from $\Delta f(\beta \to 0)$ (based on data in Fig. 3).

$^3$He. For this preplating the temperature gradient across the cavity and the bulk marker is negligible [43].

The phase diagram was mapped with NMR pulses with a small tipping angle $\beta < 10°$, applied while ramping the temperature up or down at a rate of 10–30 μK/h; see Fig. 2(a). The superfluidity manifests as an NMR precession frequency shift $\Delta f = f - f_L$ away from the Larmor frequency, with an onset at the superfluid transition temperature $T_c$. The minute suppression of $T_c$ with respect to $T_{c0}$ [Fig. 1(c)] indicates that the surface scattering is nearly perfectly specular. Within the quasiclassical theory [60,65,76] this can be parametrized in terms of specularity $S$, the probability of specular scattering [43,64]. Our data correspond to $S > 0.97$ over the entire pressure range, consistent with $T_c$ suppression measured in a $D = 192$ nm cavity [43]. The physical origin of the observed deviation from perfect specularity is not understood at this time. Candidates are nonuniformity of the adsorption potential of the silicon substrate and finite solubility of $^3$He in $^4$He [46].

The calculations using quasiclassical theory demonstrate that in the case of the $A$ phase, $S = 0.98$ surface specularity reduces the energy gap in a $D = 80$ nm cavity only by 2.5%–5.0% at 0.2 bar and by 0.5%–1.0% at 21.0 bar with respect to the bulk gap achieved for $S = 1.0$ [46]. Thus we directly use the magnitude of the measured frequency shift, proportional to the square of the energy gap, to identify the confined superfluid as the $A$ phase close to $T_c$ and to measure the bulk $A$-phase gap over a wide temperature and pressure range.

The temperature-independent magnitude of the cavity signal (constant sample magnetization), as exemplified in Fig. 2(a), indicates equal spin pairing, consistent with $^3$He-$A$. The superfluid frequency shift $\Delta f$ is negative, as expected for the dipole-unlocked spin-orbit orientation of the $A$ phase, previously identified in less confined systems [58,61,74,81]. This orientation is characterized by mutually perpendicular orbital angular momentum $\hat{\mathbf{l}}$ and zero-spin direction $\hat{\mathbf{d}}$, with the former locked perpendicular to the surface throughout the cavity by the strong confinement and the latter perpendicular to the magnetic field. At small tipping angles ($\cos\beta \approx 1$) the frequency shift in this orientation is of the same magnitude and opposite sign as in the bulk. To confirm this, the initial slope $IS_f = \partial|f^2 - f_L^2|/\partial(1 - T/T_c)$ of the measured frequency shift is in good agreement with prior A-phase measurements [Fig. 2(b)]. In this Letter we determine the initial slopes over the $0.9T_c < T < T_c$ temperature range where the frequency shift is still expected to vary linearly with temperature [43,46]. Further verification of the A phase with $\hat{\mathbf{d}} \perp \hat{\mathbf{l}} \| \mathbf{H}_0$ is the tipping-angle dependence $\Delta f(\beta) \propto -\cos\beta$ [58] [Fig. 2(c)].

We now review the NMR signatures of the planar ($^3$He-P) phase, the strongest (and, to our knowledge, the only) alternative candidate for the superfluid state in our cavity consistent with the temperature-independent magnetization. Since the planar phase is the extreme limit of the B phase with planar distortion [61,82,83], two nonequivalent spin-orbit orientations can be realized in a cavity with magnetic field normal to it. We readily rule out the stable orientation $P_+$ characterized by $\Delta f > 0$. In the metastable $P_-$ state, $\Delta f$ has the same sign and tipping-angle dependence as the dipole-unlocked A phase in our geometry. Moreover, its magnitude coincides with that of the A phase in the weak-coupling limit, making it difficult to distinguish these phases from the NMR signatures alone. However, below a critical magnetic field of order the dipole field $H_D \sim 3$ mT [35], $P_-$ becomes unstable and converts into $P_+$ [82]. Thus, we ruled out the planar phase by observing no change to the negative frequency shift upon cycling the magnetic field down to zero and back up [46].

Figure 3 shows the A-phase energy gap inferred from the small-tipping-angle frequency shift via pressure-dependent temperature-independent scaling [43]

$$\Delta_0^2(T) = \frac{IS_\Delta}{IS_f}|f^2(T) - f_L^2|. \quad (1)$$

Here, $\Delta_0$ refers to the maximum A-phase energy gap in the momentum space at $\hat{\mathbf{p}} \perp \hat{\mathbf{l}}$. The initial slope of the gap $IS_\Delta = \partial\Delta_0^2/\partial(1 - T/T_{c0}) \propto \Delta C_A \propto 1/\beta_{245}$ is directly related to the A-phase heat capacity jump $\Delta C_A$ at $T_{c0}$, which is inversely proportional to the Ginzburg-Landau parameter $\beta_{245}$ [35,84].

The data clearly deviate from the weak-coupling gap $\Delta_{0,BCS}(T)$ obtained within the p-wave BCS theory. This is a manifestation of strong-coupling effects, particularly prominent at high pressures [76]. Agreement near $T_{c0}$ can be achieved by scaling the BCS gap [43] by the ratio of the $\beta_{245}^{BCS}$ obtained from the BCS theory and $\beta_{245}$ derived experimentally at a given pressure [84],

$$\Delta_0^2(T) = \frac{\beta_{245}^{BCS}}{\beta_{245}}\Delta_{0,BCS}^2(T). \quad (2)$$

However, Eq. (2) overestimates the data at low temperatures and high pressures. This is consistent with the reduction of the strong-coupling effects on cooling, which has been incorporated into Ginzburg-Landau theory in terms of temperature-dependent $\beta_i$ parameters [85]. We introduce an ansatz for this temperature dependence [46]:

$$\beta_{245}(T) = \beta_{245}^{BCS} + (\beta_{245}(T_{c0}) - \beta_{245}^{BCS}) \times \left(1 - 0.09\frac{\Delta_{0,BCS}^2(T)}{k_B^2 T_{c0}^2}\right), \quad (3)$$

where $\beta_{245}(T_{c0}) \equiv \beta_{245}$ is the conventional Ginzburg-Landau parameter measured near $T_{c0}$ [84]. Substituting Eq. (3) into Eq. (2) leads to excellent agreement with the experimental data over the full temperature and pressure range. We propose that Eq. (3) and similar expressions for the other $\beta_i$ parameters may improve the quantitative predictions of the Ginzburg-Landau theory at low temperatures. Moreover, the measured energy gap offers a stringent test of the strong-coupling quasiclassical theory of $^3$He [86].

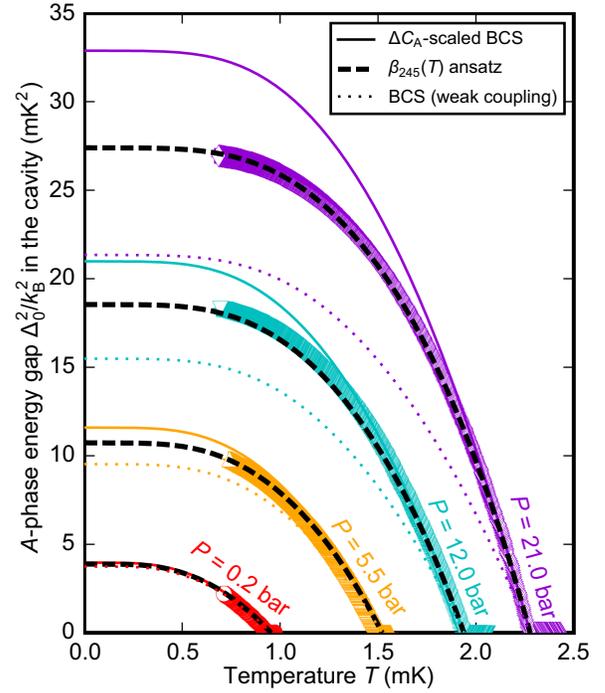

FIG. 3. The temperature dependence of the A-phase energy gap $\Delta_0^2/k_B^2$ inferred from NMR using Eq. (1) (open symbols) in comparison with various theoretical models. The pressure (color coded) tunes the strong-coupling effects. The BCS gap (dotted lines), fully determined by $T_{c0}$, undershoots the data. Near $T_{c0}$ the agreement is improved by scaling the BCS gap using Eq. (2) with experimentally determined $\beta_{245}$ [84] (solid lines). The ansatz, Eq. (3), for the temperature dependence of $\beta_{245}$ describes the measured gap over the full temperature range (dashed lines). One more pressure is shown in Supplemental Material [46].

In conclusion, the constant signal magnitude, as well as the sign, tipping-angle dependence, and the initial slope of the frequency shift $\Delta f$ very strongly suggest that the chiral $^3$He-A is the only stable superfluid phase under strong confinement down to $D/\xi_0 = 1$. The earlier observation of a discontinuity in the superfluid density as a function of film thickness at $D = 137$ nm was obtained using a torsional oscillator with a mechanically polished, but still relatively rough, Cu disk as the substrate for the $^3$He film [44]. Thus the observed transport anomaly possibly reflects a change in film morphology, with extra superflow paths in thicker films. In contrast to our experiment, the boundaries of this film were a free surface and a solid $^3$He boundary layer, which we subsequently have found to influence the surface quasiparticle scattering [43]. Determination of the order parameter in this case is a goal of future work.

The possibility of creating nearly perfectly specular surfaces by $^4$He preplating provides a mechanism to observe unsuppressed superfluidity in thin films. This will enable future investigations deeper into the two-dimensional regime $D/\xi_0 < 1$ [87]. Treating a thin helium film within the confining slab-shaped cavity as trapped in an infinite potential well in the $z$ direction, the quantization of the wave vector $k_z(n) = \pi n/D$ limits the allowed values of the $A$-phase energy gap to [46]

$$\Delta_A(n) = \Delta_0 \sqrt{1 - \frac{n^2}{n_0^2}}. \quad (4)$$

Here, $n \leq n_0$ is an integer and $n_0 = k_F D/\pi$ determines the maximum available number of Fermi disks for $0 < k_z \leq k_F$, where $k_z(n_0) = k_F$ at the pole of the Fermi surface. By fine tuning the system, using pressure to adjust the quantization condition to align the largest allowed $k_z$ away from $k_F$, one could eliminate nodes entirely from the energy gap (see Fig. 4). In our $D = 80$ nm cavity $n_0 \approx 200$, which sets the Fermi disks too densely to see any quantization effects within the temperature range covered by us. However, taking, for example, a $D = 10$ nm cavity, where $n_0 \approx 25$, the minimum value of the gap corresponding to the largest available $k_z$ could be as high as $0.28\Delta_0$ [46]. Such a fully gapped chiral superfluid at very low—but achievable—temperatures, without the usual node-bound thermal quasiparticle excitations present, enables access to the physics of the two-dimensional superfluid film [34]. We also note that as the normal state becomes progressively more 2D within more confined systems, the Fermi liquid properties and spin-fluctuation pairing interactions may be modified, and eventually impact the stability of the 2D $^3$He-$A$ as well [46]. Until that limit the quantization of $k_z$ should enhance the quest to detect the emergent chiral-specific phenomena, such as the quantum Hall effect [2,5] and the edge-bound chiral Majorana-Weyl states and the corresponding ground-state edge currents. In addition, confining such a thin $^3$He-$A$ film laterally has been predicted to lead to a sequence of phase transitions into a pair-density-wave phase and the polar phase [88].

*Acknowledgments*—We thank J. A. Sauls for helpful discussions. The research leading to these results has received funding in the UK from the UKRI EPSRC under EP/R04533X/1 and from the UKRI STFC under ST/T006749/1 ("QUEST-DMC"). J. M. P. acknowledges funding from the NSF under DMR-1202991 and DMR-2401589, and A. V. acknowledges funding from the NSF under DMR-2023928. In addition, the work has been supported by the European Union's Horizon 2020 Research and Innovation Programme under Grant agreement No. 824109 (European Microkelvin Platform). Nanofabrication was carried out at the Cornell Nanofabrication Facility (CNF) with assistance and advice from technical staff. Measurements were made at the London Low Temperature Laboratory, where we acknowledge the excellent support of technical staff, in particular Richard Elsom, Ian Higgs, Paul Bamford, and Harpal Sandhu.

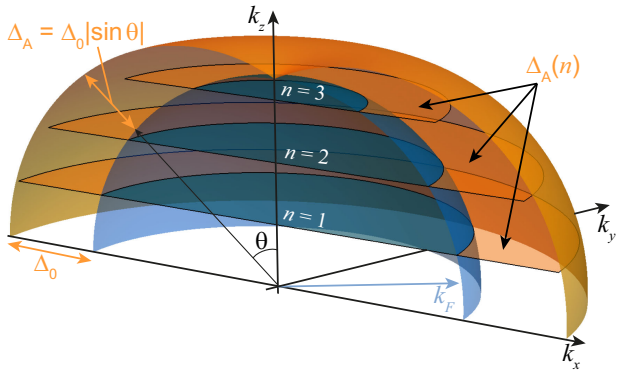

FIG. 4. Illustration of the allowed substates of the Fermi sphere and the $A$-phase energy gap resulting from the quantization of $k_z$ in a very thin film of superfluid $^3$He. The blue quarter sphere represents part of the Fermi sphere with radius $k_F$. The orange surface around it is a schematic illustration of the bulk $A$-phase energy gap $\Delta_A = \Delta_0|\sin\theta|$, where $\theta$ is the angle between the anisotropy axis $\hat{\mathbf{l}} \parallel \hat{\mathbf{z}}$ and $\hat{\mathbf{k}}$ [35]. The allowed substates of $k_z$ due to the size quantization are drawn as disks labeled by $n$. In this example the maximum number of substates of $k_z$ below $k_F$ is 3, and the highest substate does not coincide with the node in the gap, i.e., $3 < n_0 < 4$. The energy gap $\Delta_A(n)$ corresponding to each disk follows from Eq. (4). The gap and the Fermi sphere are not drawn to scale since $\Delta_0 \ll E_F$, where $E_F$ is the Fermi energy.

*Data availability*—The measured $T_c$, initial slopes, and the energy gap data and the related calculations that support the findings of this Letter are openly available in Figshare [89].

[1] C. Kallin and J. Berlinsky, Chiral superconductors, Rep. Prog. Phys. **79**, 054502 (2016).
[2] G. E. Volovik, Quantum Hall state and chiral edge state in thin $^3$He-$A$ film, JETP Lett. **55**, 368 (1992).
[3] T. Kita, Angular momentum of anisotropic superfluids at finite temperatures, J. Phys. Soc. Jpn. **67**, 216 (1998).

[4] M. Stone and R. Roy, Edge modes, edge currents, and gauge invariance in $p_x + ip_y$ superfluids and superconductors, Phys. Rev. B 69, 184511 (2004).

[5] G. E. Volovik, An analog of the quantum Hall effect in a superfluid $^3$He film, Sov. Phys. JETP 67, 1804 (1988).

[6] N. Read and D. Green, Paired states of fermions in two dimensions with breaking of parity and time-reversal symmetries and the fractional quantum Hall effect, Phys. Rev. B 61, 10267 (2000).

[7] G. E. Volovik and V. P. Mineev, Line and point singularities in superfluid $^3$He, JETP Lett. 24, 561 (1976).

[8] M. M. Salomaa and G. E. Volovik, Half-quantum vortices in superfluid $^3$He-A, Phys. Rev. Lett. 55, 1184 (1985).

[9] G. E. Volovik, Monopole, half-quantum vortex, and nexus in chiral superfluids and superconductors, JETP Lett. 70, 792 (1999).

[10] D. A. Ivanov, Non-Abelian statistics of half-quantum vortices in p-wave superconductors, Phys. Rev. Lett. 86, 268 (2001).

[11] J. A. Sauls, Surface states, edge currents, and the angular momentum of chiral p-wave superfluids, Phys. Rev. B 84, 214509 (2011).

[12] G. Moore and N. Read, Nonabelions in the fractional quantum Hall effect, Nucl. Phys. B360, 362 (1991).

[13] G. E. Volovik, Fermion zero modes on vortices in chiral superconductors, JETP Lett. 70, 609 (1999).

[14] A. P. Schnyder, S. Ryu, A. Furusaki, and A. W. W. Ludwig, Classification of topological insulators and superconductors in three spatial dimensions, Phys. Rev. B 78, 195125 (2008).

[15] M. Sato and Y. Ando, Topological superconductors: A review, Rep. Prog. Phys. 80, 076501 (2017).

[16] A. Yu. Kitaev, Fault-tolerant quantum computation by anyons, Ann. Phys. (Amsterdam) 303, 2 (2003).

[17] C. W. J. Beenakker, Search for Majorana fermions in superconductors, Annu. Rev. Condens. Matter Phys. 4, 113 (2013).

[18] S. Das Sarma, M. Freedman, and C. Nayak, Majorana zero modes and topological quantum computation, npj Quantum Inf. 1, 15001 (2015).

[19] B. Lian, X.-Q. Sun, A. Vaezi, X.-L. Qi, and S.-C. Zhang, Topological quantum computation based on chiral Majorana fermions, Proc. Natl. Acad. Sci. U.S.A. 115, 10938 (2018).

[20] S. K. Ghosh, M. Smidman, T. Shang, J. F. Annett, A. D. Hillier, J. Quintanilla, and H. Yuan, Recent progress on superconductors with time-reversal symmetry breaking, J. Phys. Condens. Matter 33, 033001 (2020).

[21] A. Ramires, Symmetry aspects of chiral superconductors, Contemp. Phys. 63, 71 (2022).

[22] A. Pustogow, Y. Luo, A. Chronister, Y.-S. Su, D. A. Sokolov, F. Jerzembeck, A. P. Mackenzie, C. W. Hicks, N. Kikugawa, S. Raghu, E. D. Bauer, and S. E. Brown, Constraints on the superconducting order parameter in $Sr_2RuO_4$ from oxygen-17 nuclear magnetic resonance, Nature (London) 574, 72 (2019).

[23] H. G. Suh, H. Menke, P. M. R. Brydon, C. Timm, A. Ramires, and D. F. Agterberg, Stabilizing even-parity chiral superconductivity in $Sr_2RuO_4$, Phys. Rev. Res. 2, 032023(R) (2020).

[24] V. Grinenko, D. Das, R. Gupta, B. Zinkl, N. Kikugawa, Y. Maeno, C. W. Hicks, H.-H. Klauss, M. Sigrist, and R. Khasanov, Unsplit superconducting and time reversal symmetry breaking transitions in $Sr_2RuO_4$ under hydrostatic pressure and disorder, Nat. Commun. 12, 3920 (2021).

[25] R. Joynt and L. Taillefer, The superconducting phases of $UPt_3$, Rev. Mod. Phys. 74, 235 (2002).

[26] J. D. Strand, D. J. Van Harlingen, J. B. Kycia, and W. P. Halperin, Evidence for complex superconducting order parameter symmetry in the low-temperature phase of $UPt_3$ from Josephson interferometry, Phys. Rev. Lett. 103, 197002 (2009).

[27] E. R. Schemm, W. J. Gannon, C. M. Wishne, W. P. Halperin, and A. Kapitulnik, Observation of broken time-reversal symmetry in the heavy-fermion superconductor $UPt_3$, Science 345, 190 (2014).

[28] K. E. Avers, W. J. Gannon, S. J. Kuhn, W. P. Halperin, J. A. Sauls, L. DeBeer-Schmitt, C. D. Dewhurst, J. Gavilano, G. Nagy, U. Gasser, and M. R. Eskildsen, Broken time-reversal symmetry in the topological superconductor $UPt_3$, Nat. Phys. 16, 531 (2020).

[29] L. Jiao, S. Howard, S. Ran, Z. Wang, J. O. Rodriguez, M. Sigrist, Z. Wang, N. P. Butch, and V. Madhavan, Chiral superconductivity in heavy-fermion metal $UTe_2$, Nature (London) 579, 523 (2020).

[30] P. K. Biswas, S. K. Ghosh, J. Z. Zhao, D. A. Mayoh, N. D. Zhigadlo, X. Xu, C. Baines, A. D. Hillier, G. Balakrishnan, and M. R. Lees, Chiral singlet superconductivity in the weakly correlated metal $LaPt_3P$, Nat. Commun. 12, 2504 (2021).

[31] P. K. Biswas, H. Luetkens, T. Neupert, T. Stürzer, C. Baines, G. Pascua, A. P. Schnyder, M. H. Fischer, J. Goryo, M. R. Lees, H. Maeter, F. Brückner, H.-H. Klauss, M. Nicklas, P. J. Baker, A. D. Hillier, M. Sigrist, A. Amato, and D. Johrendt, Evidence for superconductivity with broken time-reversal symmetry in locally noncentrosymmetric SrPtAs, Phys. Rev. B 87, 180503(R) (2013).

[32] M. H. Fischer, T. Neupert, C. Platt, A. P. Schnyder, W. Hanke, J. Goryo, R. Thomale, and M. Sigrist, Chiral d-wave superconductivity in SrPtAs, Phys. Rev. B 89, 020509(R) (2014).

[33] J. A. Mydosh, P. M. Oppeneer, and P. Riseborough, Hidden order and beyond: An experimental—theoretical overview of the multifaceted behavior of $URu_2Si_2$, J. Phys. Condens. Matter 32, 143002 (2020).

[34] G. E. Volovik, *Exotic Properties of Superfluid Helium 3*, Series in Modern Condensed Matter Physics Vol. 1 (World Scientific, Singapore, 1992).

[35] D. Vollhardt and P. Wölfle, *The Superfluid Phases of Helium 3*, Dover ed. (Dover Publications, New York, 2013).

[36] G. E. Volovik, *The Universe in a Helium Droplet*, International Series of Monographs on Physics Vol. 117 (Oxford University Press, Oxford, 2003).

[37] P. M. Walmsley and A. I. Golov, Chirality of superfluid $^3$He-A, Phys. Rev. Lett. 109, 215301 (2012).

[38] H. Ikegami, Y. Tsutsumi, and K. Kono, Chiral symmetry breaking in superfluid $^3$He-A, Science 341, 59 (2013).

[39] H. Ikegami, Y. Tsutsumi, and K. Kono, Observation of intrinsic magnus force and direct detection of chirality in superfluid $^3$He-A, J. Phys. Soc. Jpn. 84, 044602 (2015).

[40] O. Shevtsov and J. A. Sauls, Electron bubbles and Weyl fermions in chiral superfluid $^3$He-A, Phys. Rev. B **94,** 064511 (2016).

[41] J. Kasai, Y. Okamoto, K. Nishioka, T. Takagi, and Y. Sasaki, Chiral domain structure in superfluid $^3$He-A studied by magnetic resonance imaging, Phys. Rev. Lett. **120,** 205301 (2018).

[42] T. Mizushima, Y. Tsutsumi, M. Sato, and K. Machida, Symmetry protected topological superfluid $^3$He-B, J. Phys. Condens. Matter **27,** 113203 (2015).

[43] P. J. Heikkinen, A. Casey, L. V. Levitin, X. Rojas, A. Vorontsov, P. Sharma, N. Zhelev, J. M. Parpia, and J. Saunders, Fragility of surface states in topological superfluid $^3$He, Nat. Commun. **12,** 1574 (2021).

[44] J. Xu and B. C. Crooker, Very thin films of $^3$He: A new phase?, Phys. Rev. Lett. **65,** 3005 (1990).

[45] We note that Ref. [44] reports a potential phase transition between 273 and 277 nm. As discussed therein, these quoted values are twice the actual measured film thickness to map unsaturated films grown on a solid substrate to slabs sandwiched between two solid walls, taking into account the expected full specularity of the free surface. See also Ref. [46].

[46] See Supplemental Material at http://link.aps.org/supplemental/10.1103/PhysRevLett.134.136001 for the comparison between unsaturated films and confined helium, the near-specular boundary conditions and related gap suppression, a discussion of initial slopes, a derivation of the strong-coupling model, the ruling out of the planar phase, and details on the size quantization of the energy gap, which includes Refs. [47–56].

[47] J. G. Daunt, R. F. Harris-Lowe, J. P. Harrison, A. Sachrajda, S. Steel, R. R. Turkington, and P. Zawadski, Critical temperature and critical current of thin-film superfluid $^3$He, J. Low Temp. Phys. **70,** 547 (1988).

[48] M. Saitoh, H. Ikegami, and K. Kono, Onset of superfluidity in $^3$He films, Phys. Rev. Lett. **117,** 205302 (2016).

[49] H. Ikegami and K. Kono, Review: Observation of Majorana bound states at a free surface of $^3$He-B, J. Low Temp. Phys. **195,** 343 (2019).

[50] B. Ilic (private communication).

[51] S. M. Tholen and J. M. Parpia, Effect of $^4$He on the surface scattering of $^3$He, Phys. Rev. B **47,** 319 (1993).

[52] S. Murakawa, M. Wasai, K. Akiyama, Y. Wada, Y. Tamura, R. Nomura, and Y. Okuda, Strong suppression of the Kosterlitz-Thouless transition in a $^4$He film under high pressure, Phys. Rev. Lett. **108,** 025302 (2012).

[53] Y. Okuda and R. Nomura, Surface Andreev bound states of superfluid $^3$He and Majorana fermions, J. Phys. Condens. Matter **24,** 343201 (2012).

[54] J. Saunders, B. Cowan, and J. Nyéki, Atomically layered helium films at ultralow temperatures: Model systems for realizing quantum materials, J. Low Temp. Phys. **201,** 615 (2020).

[55] A. Waterworth, A SQUID NMR study of $^3$He-$^4$He mixture films adsorbed on graphite, Ph.D. thesis, Royal Holloway, University of London, 2019.

[56] P. Sharma, A. Córcoles, R. G. Bennett, J. M. Parpia, B. Cowan, A. Casey, and J. Saunders, Quantum transport in mesoscopic $^3$He films: Experimental study of the interference of bulk and boundary scattering, Phys. Rev. Lett. **107,** 196805 (2011).

[57] A. L. Fetter and S. Ullah, Superfluid density and critical current of $^3$He in confined geometries, J. Low Temp. Phys. **70,** 515 (1988).

[58] M. R. Freeman, R. S. Germain, E. V. Thuneberg, and R. C. Richardson, Size effects in thin films of superfluid $^3$He, Phys. Rev. Lett. **60,** 596 (1988).

[59] Y. Nagato and K. Nagai, A–B transition of superfluid $^3$He in a slab with rough surfaces, Physica (Amsterdam) **284B,** 269 (2000).

[60] A. B. Vorontsov and J. A. Sauls, Thermodynamic properties of thin films of superfluid $^3$He-A, Phys. Rev. B **68,** 064508 (2003).

[61] L. V. Levitin, R. G. Bennett, A. Casey, B. Cowan, J. Saunders, D. Drung, Th. Schurig, and J. M. Parpia, Phase diagram of the topological superfluid $^3$He confined in a nanoscale slab geometry, Science **340,** 841 (2013).

[62] V. Ambegaokar, P. G. deGennes, and D. Rainer, Landau-Ginsburg equations for an anisotropic superfluid, Phys. Rev. A **9,** 2676 (1974).

[63] L. H. Kjäldman, J. Kurkijärvi, and D. Rainer, Suppression of P-wave superfluidity in long, narrow pores, J. Low Temp. Phys. **33,** 577 (1978).

[64] Y. Nagato, M. Yamamoto, and K. Nagai, Rough surface effects on the p-wave Fermi superfluids, J. Low Temp. Phys. **110,** 1135 (1998).

[65] A. B. Vorontsov, Andreev bound states in superconducting films and confined superfluid $^3$He, Phil. Trans. R. Soc. A **376,** 20150144 (2018).

[66] L. V. Levitin, R. G. Bennett, A. Casey, B. P. Cowan, C. P. Lusher, J. Saunders, D. Drung, and Th. Schurig, A nuclear magnetic resonance spectrometer for operation around 1 MHz with a sub-10-mK noise temperature, based on a two-stage dc superconducting quantum interference device sensor, Appl. Phys. Lett. **91,** 262507 (2007).

[67] P. J. Heikkinen, N. Eng, L. V. Levitin, X. Rojas, A. Singh, S. Autti, R. P. Haley, M. Hindmarsh, D. E. Zmeev, J. M. Parpia, A. Casey, and J. Saunders, Nanofluidic platform for studying the first-order phase transitions in superfluid helium-3, J. Low Temp. Phys. **215,** 477 (2024).

[68] S. Dimov, R. G. Bennett, A. Córcoles, L. V. Levitin, B. Ilic, S. S. Verbridge, J. Saunders, A. Casey, and J. M. Parpia, Anodically bonded submicron microfluidic chambers, Rev. Sci. Instrum. **81,** 013907 (2010).

[69] N. Zhelev, T. S. Abhilash, R. G. Bennett, E. N. Smith, B. Ilic, J. M. Parpia, L. V. Levitin, X. Rojas, A. Casey, and J. Saunders, Fabrication of microfluidic cavities using Si-to-glass anodic bonding, Rev. Sci. Instrum. **89,** 073902 (2018).

[70] R. S. Bonilla, B. Hoex, P. Hamer, and P. R. Wilshaw, Dielectric surface passivation for silicon solar cells: A review, Phys. Status Solidi A **214,** 1700293 (2017).

[71] R. Kotipalli, R. Delamare, O. Poncelet, X. Tang, L. A. Francis, and D. Flandre, Passivation effects of atomic-layer-deposited aluminum oxide, EPJ Photovoltaics **4,** 45107 (2013).

[72] N. Eng, Development of next-generation scientific platforms for moonshot quantum technologies research, Ph.D. thesis, Royal Holloway, University of London, 2024.

[73] D. S. Greywall, ³He specific heat and thermometry at millikelvin temperatures, Phys. Rev. B **33,** 7520 (1986).

[74] M. R. Freeman and R. C. Richardson, Size effects in superfluid ³He films, Phys. Rev. B **41,** 11011 (1990).

[75] S. M. Tholen and J. M. Parpia, Slip and the effect of ⁴He at the ³He-silicon interface, Phys. Rev. Lett. **67,** 334 (1991).

[76] J. W. Serene and D. Rainer, The quasiclassical approach to superfluid ³He, Phys. Rep. **101,** 221 (1983).

[77] M. R. Rand, Nonlinear spin dynamics and magnetic field distortion of the superfluid ³He-B order parameter, Ph.D. thesis, Northwestern University, 1996.

[78] M. R. Rand, H. H. Hensley, J. B. Kycia, T. M. Haard, Y. Lee, P. J. Hamot, and W. P. Halperin, New NMR evidence: Can an axi-planar superfluid ³He-A order parameter be ruled out?, Physica (Amsterdam) **194-196B,** 805 (1994).

[79] A. M. Zimmerman, M. D. Nguyen, and W. P. Halperin, NMR frequency shifts and phase identification in superfluid ³He, J. Low Temp. Phys. **195,** 358 (2019).

[80] P. Schiffer, M. T. O'Keefe, H. Fukuyama, and D. D. Osheroff, Low-temperature studies of the NMR frequency shift in superfluid ³He-A, Phys. Rev. Lett. **69,** 3096 (1992).

[81] A. I. Ahonen, M. Krusius, and M. A. Paalanen, NMR experiments on the superfluid phases of ³He in restricted geometries, J. Low Temp. Phys. **25,** 421 (1976).

[82] L. V. Levitin, R. G. Bennett, E. V. Surovtsev, J. M. Parpia, B. Cowan, A. J. Casey, and J. Saunders, Surface-induced order parameter distortion in superfluid ³He-B measured by nonlinear NMR, Phys. Rev. Lett. **111,** 235304 (2013).

[83] L. V. Levitin, B. Yager, L. Sumner, B. Cowan, A. J. Casey, J. Saunders, N. Zhelev, R. G. Bennett, and J. M. Parpia, Evidence for a spatially modulated superfluid phase of ³He under confinement, Phys. Rev. Lett. **122,** 085301 (2019).

[84] H. Choi, J. P. Davis, J. Pollanen, T. M. Haard, and W. P. Halperin, Strong coupling corrections to the Ginzburg-Landau theory of superfluid ³He, Phys. Rev. B **75,** 174503 (2007).

[85] J. J. Wiman and J. A. Sauls, Superfluid phases of ³He in nanoscale channels, Phys. Rev. B **92,** 144515 (2015).

[86] J. J. Wiman, Quantitative superfluid helium-3 from confinement to bulk, Ph.D. thesis, Northwestern University, 2018.

[87] J. Saunders, Realizing quantum materials with helium: Helium films at ultralow temperatures, from strongly correlated atomically layered films to topological superfluidity, in *Topological Phase Transitions and New Developments*, edited by L. Brink, M. Gunn, J. V. Jose, J. M. Kosterlitz, and K. K. Phua (World Scientific Publishing, Singapore, 2019), pp. 165–196.

[88] H. Wu and J. A. Sauls, Weyl Fermions and broken symmetry phases of laterally confined ³He films, J. Phys. Condens. Matter **35,** 495402 (2023).

[89] P. J. Heikkinen, L. V. Levitin, X. Rojas, A. Singh, N. Eng, A. Vorontsov, N. Zhelev, T. S. Abhilash, J. M. Parpia, A. Casey, and J. Saunders, Accompanying data for "Chiral superfluid helium-3 in the quasi-two-dimensional limit" (2024), Figshare, 10.17637/rh.27020611.